\documentclass[fleqn,usenatbib]{mnras}

\usepackage[T2A,T1]{fontenc}
\usepackage[utf8]{inputenc}
\usepackage{ae,aecompl}

\usepackage{graphicx}	
\usepackage{amsmath}	
\usepackage{amssymb}	

\title[Relativistic polytrope]{Relativistic polytrope from the collimation and acceleration profiles of the M87 jet at subparsec scales\\and thermodynamic evidence for the Blandford-Znajek mechanism}

\author[D. N. Sob'yanin]{
Denis Nikolaevich Sob'yanin\thanks{E-mail: sobyanin@lpi.ru}
\fontencoding{T2A}\selectfont
 (Денис Николаевич Собьянин)
\fontencoding{T1}\selectfont
\\
I. E. Tamm Division of Theoretical Physics, P. N. Lebedev Physical Institute of the Russian Academy of Sciences,\\Leninskii Prospekt 53, Moscow 119991, Russia
}

\date{Accepted 2019 July 17. Received 2019 July 13; in original form 2019 May 28}

\pubyear{2019}
\volume{489}
\pagerange{L7--L11}

\begin{document}
\label{firstpage}
\maketitle

\begin{abstract}
Recent Very Long Baseline Interferometry observations of the relativistic jet in the M87 radio galaxy at 43~GHz show gradual relativistic acceleration of the plasma and suggest a linear dependence of Lorentz factor on jet radius at scales up to 8~marcsec (0.65~pc) from the core (2.5~marcsec in projection). General analysis of integrals of motion being unaltered along the jet and reflecting fundamental conservation laws shows that the above dependence implies a polytropic equation of state with index 4/3. The inferred value of the polytropic index appears independent of the exact nature of forces sustaining the transverse balance of the jet and indicates exact conservation of the longitudinal electric current and hence the existence of a stable internal electromagnetic structure at the scales under consideration. At this index the flow is hot and corresponds to relativistic thermodynamic motion of particles. Considerable weakening of the acceleration efficiency after 8~marcsec with the jet form being unchanged can be related to the plasma cooling, when the polytropic index becomes 5/3. Such a sharp change in the index without intermediate delay at 1.44 during cooling favours the existence of an electron-positron plasma and requires at least partial participation of the Blandford-Znajek mechanism in the launching of the M87 jet.
\end{abstract}



\section{Introduction}

Today, the nearby giant elliptical galaxy M87 attracts significant attention as a host for a supermassive black hole with the first shadow imaged \citep{EHTM8712019}. The main reasons of the success are the use of mm wavelengths, for which source-intrinsic absorption effects reduce and the medium becomes transparent \citep{HadaEtal2011}; global Very Long Baseline Interferometry (VLBI) with the Event Horizon Telescope at such wavelengths, which allowed one to reach an ultrahigh angular resolution of $\sim20-25\text{ }\mu$arcsec; and a high ratio of black-hole mass to distance to us, so that it is possible to resolve distances in the order of several Schwarzschild radii.

Meanwhile, M87 also hosts the prominent relativistic jet known already more than $100$~yr \citep{Curtis1918}. The jet reveals itself in multifarious emission throughout the spectrum from radio to TeV, but radio is of special significance because the VLBI radio imaging for the mentioned reasons is one of the most important techniques giving us an opportunity to study fine structure of the jet, its relativistic kinematics, and long-term dynamics \citep{KovalevEtal2007,AsadaEtal2014,HadaEtal2017,WalkerEtal2018}.

The form of the jet in M87, the so-called collimation profile giving the dependence of jet radius $r$ on distance $z$ from the base, is well described by a power law
\begin{equation}
\label{powerLawForm}
r\propto z^\alpha,
\end{equation}
where the index $\alpha$ lies in the range $0.5-0.6$  \citep{AsadaNakamura2012,HadaEtal2013,KimM87Etal2016,MertensEtal2016} and corresponds to a gradual decrease, on average, in the local opening angle, which constitutes the effect of collimation and takes place at subpc to hundred-pc scales. The parabolic form remains up to deprojected angular distances $z\sim10^3$~marcsec from the base ($1\text{ marcsec}\approx0.08$~pc), when the collimation stops and a transition to a conical shape $r\propto z$ occurs \citep{AsadaNakamura2012}. Note that gradual power-law expansion occurs on average, as a sequence of subsequent expansions and contractions is possible. According to recent observations by \citet{WalkerEtal2018}, the M87 jet demonstrates three consecutive expansion/recollimation cycles up to $z_\text{obs}\sim7.4$~marcsec.

The relativistic acceleration profile for the M87 jet, which gives the dependence of Lorentz factor of the plasma on distance from the base, is measured by \citet{MertensEtal2016} and given by a power law
\begin{equation}
\label{powerLawAcceleration}
\gamma\propto r\propto z^{0.58},\text{ }z<8\text{ marcsec}.
\end{equation}
This dependence changes as deprojected distance exceeds $z\sim8$~marcsec from the base, which corresponds to projected distance $z_\text{obs}\sim2.5$~marcsec, and shows that the efficiency of acceleration significantly drops,
\begin{equation}
\label{powerLawAccelerationLargeZ}
\gamma\propto z^{0.16},\text{ }z>8\text{ marcsec}.
\end{equation}
Interestingly, the change in the $z$ dependence of Lorentz factor is not accompanied by any change in the jet form.

In this Letter, I will demonstrate that knowledge of the collimation and acceleration profiles for the M87 jet allows us to conclude about jet thermodynamics. Under general conservation laws and infinite conductivity, I will show that a polytropic equation of state for a plasma at subpc scales is relativistic and is characterized by index $4/3$. Surprisingly, this conclusion appears insensitive to specific assumptions about the nature of forces balancing the jet. The change of the acceleration profile at $z\sim8$~marcsec can be caused by the plasma cooling, and the character of the change can reflect the work of the Blandford-Znajek jet-launching mechanism.

\section{Integrals of motion}

A jet represents a flow of relativistic plasma generally governed by the Maxwell equations describing evolution of electromagnetic fields and its connection with charges and currents, by general conservation laws involving both the matter and the fields, and by some additional constraints such as high conductivity and equation of state. Consideration of the behaviour of a stationary axisymmetric ideal relativistic plasma flow simplifies greatly because we have then a jet divided by a continuum of fixed embedded magnetic tubes along which conservation of several physical quantities called integrals of motion takes place.

The simplest integral is the magnetic flux in the tube, and the value of the flux enumerates different tubes. The condition of infinite conductivity, or the freezing-in condition ($c=1$ throughout the Letter)
\begin{equation}
\label{forceFree}
\mathbf{E}=-\mathbf{v}\times\mathbf{B},
\end{equation}
where $\mathbf{E}$ and $\mathbf{B}$ are the electric and magnetic fields and $\mathbf{v}$ is the plasma velocity at a given point, with taking account of the above flux integral, transforms into conservation of the so-called Ferraro isorotation frequency
\begin{equation}
\label{GSOmegaF}
\Omega_\text{F}=\frac{v_\phi-v_\text{p}B_\phi/B_\text{p}}{r},
\end{equation}
where $v_\phi$, $v_\text{p}$ and $B_\phi$, $B_\text{p}$ are the toroidal and poloidal components of velocity and magnetic field, respectively.

The other integrals result from fundamental conservation laws combined with magnetic flux conservation. Specifically, matter conservation
\begin{equation}
\label{matterConservation}
\frac{\partial\gamma\rho}{\partial t}+\operatorname{div}\gamma\rho\mathbf{v}=0,
\end{equation}
where $\gamma=(1-v^2)^{-1/2}$ is the Lorentz factor and $\rho$ is the mass density in the comoving frame, implies conservation of
\begin{equation}
\label{GSeta}
\eta=\frac{\gamma\rho v_\text{p}}{B_\text{p}}
\end{equation}
along the magnetic tube; energy conservation
\begin{equation}
\label{energyConservation}
\frac{\partial}{\partial t}\biggl(\gamma^2 \rho h-p+\frac{E^2+B^2}{8\pi}\biggr)+\operatorname{div}\biggl(\gamma^2 \rho h\mathbf{v}+\frac{\mathbf{E}\times\mathbf{B}}{4\pi}\biggr)=0,
\end{equation}
where
\begin{equation}
\label{relativisticEnthalpy}
h=1+\varepsilon+\frac p\rho
\end{equation}
is the specific relativistic enthalpy, $\varepsilon$ is the specific internal energy, and $p$ is the pressure, requires conservation of
\begin{equation}
\label{GSE}
\mathcal{E}=\gamma h\eta-\frac{\Omega_\text{F}I}{2\pi},
\end{equation}
with $I$ being the electric current in the magnetic tube flowing through a given cross-section; while momentum conservation
\begin{equation}
\label{momentumConservation}
\begin{split}
&\frac{\partial}{\partial t}\biggl(\gamma^2 \rho h\mathbf{v}+\frac{\mathbf{E}\times\mathbf{B}}{4\pi}\biggr)\\
&+\operatorname{div}\biggl[\biggl(p+\frac{E^2+B^2}{8\pi}\biggr)\mathbf{I}+\gamma^2 \rho h\mathbf{v}\mathbf{v}-\frac{\mathbf{E}\mathbf{E}+\mathbf{B}\mathbf{B}}{4\pi}\biggr]=0,
\end{split}
\end{equation}
where $\mathbf{a}\mathbf{a}=||a_ia_j||$ is the dyad, $\mathbf{I}=||\delta_{ij}||$ is the unit tensor, and $\operatorname{div}\mathbf{T}=\nabla\cdot\mathbf{T}=||\partial T_{ji}/\partial x_j||$ is the divergence of a tensor $\mathbf{T}=||T_{ij}||$, necessitates conservation of
\begin{equation}
\label{GSL}
\mathcal{L}=\gamma h\eta r v_\phi-\frac{I}{2\pi},
\end{equation}
where $r$ is the radius of the tube at a given level over the base of the jet.

The system is complemented by an equation of state $p=p(\rho,\varepsilon)$. When the case of a polytrope with index $\Gamma$ is considered, which corresponds to
\begin{equation}
\label{EoSpolytrope}
p=(\Gamma-1)\rho\varepsilon,
\end{equation}
we have the last thermodynamic integral related to entropy,
\begin{equation}
\label{GSS}
S=\frac{p}{\rho^\Gamma},
\end{equation}
which reflects entropy conservation
\begin{equation}
\label{entropyConservation}
\frac{\partial\gamma\rho S}{\partial t}+\operatorname{div}\gamma\rho S\mathbf{v}=0.
\end{equation}
In this case the enthalpy \eqref{relativisticEnthalpy} becomes
\begin{equation}
\label{polytropeEnthalpy}
h=1+\frac\Gamma{\Gamma-1}S\rho^{\Gamma-1}.
\end{equation}

\section{Pressure and polytropic index}

\subsection{Force balance}

To find the polytropic index $\Gamma$ in the equation of state, we should realize what determines the value of pressure $p$. The radial equilibrium of the jet can be described either by a so-called Grad-Shafranov equation \citep{ChiuehEtal1991,Fendt1997,BeskinNokhrina2009} or by directly using the momentum conservation law \eqref{momentumConservation}, which in cylindrical coordinates has the form \citep{Sobyanin2017}
\begin{equation}
\label{momentumConservation2}
\biggl(p+\frac{B_z^2+B_\phi^2-E_r^2}{8\pi}\biggr)'+\frac{B_\phi^2-E_r^2}{4\pi r}-\gamma^2 \rho h\frac{v_\phi^2}{r}=0,
\end{equation}
where prime denotes the $r$ derivative. From this equation three limiting equilibrium regimes are naturally singled out that correspond to different relations of the thermodynamic pressure with other physical quantities.

\subsection{Longitudinal magnetic pressure}

The first case corresponds to the thermodynamic pressure balanced mainly by the pressure of the longitudinal magnetic field,
\begin{equation}
\label{longitudinalMagneticPressureBalance}
p\sim\frac{B_z^2}{8\pi}.
\end{equation}
Magnetic flux conservation $\Phi\sim B_z \pi r^2=\text{const}$ requires
\begin{equation}
\label{BzOnR}
B_z\propto r^{-2}
\end{equation}
and then gives the dependence of pressure on jet radius at a given level over the base,
\begin{equation}
\label{pOnR1}
p\propto r^{-4}.
\end{equation}

Conservation of the integral $\eta$ \eqref{GSeta} in the case of relativistic longitudinal motion, when $v_\text{p}\approx1$, gives the dependence of density on Lorentz factor and radius,
\begin{equation}
\label{rhoOnGammaAndR}
\rho\propto\gamma^{-1}r^{-2}.
\end{equation}
If we take for the Lorentz factor a general power-law dependence on jet radius with an index $\beta$,
\begin{equation}
\label{gammaOnRGeneral}
\gamma\propto r^\beta,
\end{equation}
we arrive at
\begin{equation}
\label{rhoOnR}
\rho\propto r^{-2-\beta}.
\end{equation}

Entropy conservation \eqref{GSS} with Eqs.~\eqref{pOnR1} and \eqref{rhoOnR} then requires a constant $r^{-4+(2+\beta)\Gamma}$, so the polytropic index for the case of a dominating longitudinal magnetic pressure is
\begin{equation}
\label{Gamma1General}
\Gamma_\text{LM}=\frac4{2+\beta}.
\end{equation}
Since from Eq.~\eqref{powerLawAcceleration} the observational index
\begin{equation}
\label{betaOne}
\beta_\text{obs}=1,\text{ }z<8\text{ marcsec},
\end{equation}
we get
\begin{equation}
\label{Gamma1}
\Gamma_\text{LM}=\frac43.
\end{equation}

\subsection{Transverse electromagnetic pressure}

Now let us consider the case when the thermodynamic pressure is balanced mainly by the forces resulting from the effect of azimuthal magnetic field and radial electric field,
\begin{equation}
\label{transverseElectromagneticPressureBalance}
p\sim\frac{B_\phi^2-E_r^2}{8\pi}.
\end{equation}
It is worth noting that the combination $B_\phi^2-E_r^2$ enters the momentum conservation law \eqref{momentumConservation2} as a unified parameter and should not be divided. The first term accounts for the pressure of toroidal magnetic field while the second (with minus sign) for the tension of radial electric field lines, and the overall effect will loosely be called a `transverse electromagnetic pressure.'

As follows from Eqs.~\eqref{forceFree} and~\eqref{GSOmegaF}, the radial electric field is
\begin{equation}
\label{Er}
E_r=-\Omega_\text{F}rB_z,
\end{equation}
therefore, since $\Omega_\text{F}$ is an integral of motion,
\begin{equation}
\label{ErOnR}
E_r\propto r^{-1}.
\end{equation}

The toroidal magnetic field is in turn expressed via the longitudinal electric current $I$,
\begin{equation}
\label{Bphi}
B_\phi=\frac{2I}r.
\end{equation}
One might think that $B_\phi\propto r^{-1}$ and hence, together with Eq.~\eqref{ErOnR}, $p\propto r^{-2}$. Importantly, this is not so. The current $I$ is generally not an exact integral of motion, and some transverse current through the magnetic surface is potentially possible. However, we will see below, Eq.~\eqref{longitudinalCurrentConservation}, that in our case $I$ is strictly conserved. The answer to the question why $p\propto r^{-2}$ does not hold is that the terms $B_\phi^2$ and $E_r^2$ almost compensate each other and the contribution to the pressure is a term of smaller order of magnitude.

Specifically, ideal conductivity \eqref{forceFree} implies
\begin{equation}
\label{ErExtended}
E_r=v_z B_\phi-v_\phi B_z.
\end{equation}
Theoretical estimations show that $B_\phi\gg B_z$ and $v_\phi\ll1$ \citep{Sobyanin2017}, so the last term in Eq.~\eqref{ErExtended} may be neglected,
\begin{equation}
\label{ErApprox}
E_r\approx v_z B_\phi.
\end{equation}
Since $v_z\approx1$, $E_r\approx B_\phi$ and the two first-order terms in the pressure are cancelled. We then obtain from Eqs.~\eqref{transverseElectromagneticPressureBalance} and \eqref{ErApprox}
\begin{equation}
\label{pOnErAndGamma}
p\sim\frac{E_r^2}{8\pi\gamma^2}.
\end{equation}
The dependence of pressure on jet radius becomes
\begin{equation}
\label{pOnR2}
p\propto r^{-2-2\beta}.
\end{equation}

Now we get from Eqs.~\eqref{GSS}, \eqref{rhoOnR}, and \eqref{pOnR2} the necessity of a constant $r^{-2(1+\beta)+(2+\beta)\Gamma}$, and the polytropic index for the case of a dominating transverse electromagnetic pressure is hence
\begin{equation}
\label{Gamma2General}
\Gamma_\text{TEM}=2\,\frac{1+\beta}{2+\beta}.
\end{equation}
Though the functional form of the dependence of $\Gamma$ on $\beta$ differs from the previous form \eqref{Gamma1General}, for the observational index \eqref{betaOne} we have again
\begin{equation}
\label{Gamma2}
\Gamma_\text{TEM}=\frac43.
\end{equation}

\subsection{Centrifugal pressure}

The last case corresponds to the pressure balanced mainly by the centrifugal pressure,
\begin{equation}
\label{centrifugalPressureBalance}
p\sim\gamma^2\rho h v_\phi^2.
\end{equation}
Here it is convenient to separately consider two subcases, the case of a cold flow, when $h\approx1$, and the case of a hot flow, when $h\gg1$. Before so doing, let us notice conservation of the quantity
\begin{equation}
\label{l}
l=\gamma h(1-\Omega_\text{F}rv_\phi),
\end{equation}
which follows from a combination of the integrals \eqref{GSeta}, \eqref{GSE}, and~\eqref{GSL}.

\subsubsection{Cold flow}

In the first case we have
\begin{equation}
\label{centrifugalPressureBalanceCold}
p\sim\gamma^2\rho v_\phi^2.
\end{equation}
Taking into account non-relativistic motion at the jet base, we have the equality
\begin{equation}
\label{lNonrelCold}
1\approx\gamma(1-\Omega_\text{F}rv_\phi),
\end{equation}
which was utilized earlier to estimate the Ferraro isorotation frequency for the M87 jet \citep{MertensEtal2016}. For large Lorentz factors it transforms to
\begin{equation}
\label{lNonrelColdLargeLorentz}
\Omega_\text{F}rv_\phi\sim1,
\end{equation}
so that
\begin{equation}
\label{vPhiOnR}
v_\phi\propto r^{-1}.
\end{equation}
Combining Eqs.~\eqref{gammaOnRGeneral}, \eqref{rhoOnR}, \eqref{centrifugalPressureBalanceCold}, and~\eqref{vPhiOnR}, we conclude
\begin{equation}
\label{pOnR3Cold}
p\propto r^{\beta-4}.
\end{equation}

Remembering entropy conservation~\eqref{GSS}, we need a constant $r^{\beta-4+(2+\beta)\Gamma}$, which implies the polytropic index
\begin{equation}
\label{Gamma3ColdGeneral}
\Gamma_\text{CC}=\frac{4-\beta}{2+\beta},
\end{equation}
for the observed $\beta$ being
\begin{equation}
\label{Gamma3Cold}
\Gamma_\text{CC}=1.
\end{equation}
We see an impossible value of the index $\Gamma_\text{CC}<4/3$, which means that the case of a cold flow with centrifugal pressure is not realized for the M87 jet at the distances considered.

\subsubsection{Hot flow}

In the second case we may write from Eq.~\eqref{polytropeEnthalpy}
\begin{equation}
\label{polytropeEnthalpyHot}
h\propto\rho^{\Gamma-1}.
\end{equation}
Since $\rho h\propto\rho^\Gamma$, entropy conservation requires, see Eq.~\eqref{centrifugalPressureBalance},
\begin{equation}
\label{gammaVphiConst}
\gamma v_\phi=\text{const}.
\end{equation}
Let us immediately adopt the observed value $\beta=1$; Eq.~\eqref{gammaVphiConst} becomes
\begin{equation}
\label{rVphiConst}
r v_\phi=\text{const}.
\end{equation}
Then we conclude from Eq.~\eqref{l} that $\gamma h$ is an exact integral of motion,
\begin{equation}
\label{gammaHConst}
\gamma h=\text{const},
\end{equation}
which implies
\begin{equation}
\label{centrifugalPressureBalanceHot}
p\propto\rho h\propto\frac\rho\gamma.
\end{equation}

The dependence of pressure on jet radius coincides formally with that for the case of a dominating transverse electromagnetic pressure,
\begin{equation}
\label{pOnR3Hot}
p\propto r^{-2-2\beta}\propto r^{-4},
\end{equation}
so we again have the polytropic index
\begin{equation}
\label{Gamma3Hot}
\Gamma_\text{CH}=\frac43.
\end{equation}

Note for completeness that one may use the formula analogous to Eq.~\eqref{Gamma2General} for $\beta\neq1$,
\begin{equation}
\label{Gamma3HotGeneral}
\Gamma_\text{CH}=2\,\frac{1+\beta}{2+\beta},
\end{equation}
but one should then verify that the equality \eqref{gammaHConst} holds at least approximately. This will be the case when e.g. $\Omega_\text{F}rv_\phi\ll1$ (see Eq.~\eqref{l}) or when $I\approx\text{const}$ or $I=0$ (see Eq.~\eqref{GSE}).

\section{Conclusion and discussion}

We have obtained the same polytropic index $\Gamma=4/3$ irrespective of assumptions about the exact character of forces making the main contribution to balancing the thermodynamic pressure and thus providing the transverse jet equilibrium; at this index, a linear combination of forces of different nature, say, with comparable contributions to the net equilibrium is also possible. This means that the jet is hot at subparsec scales, so that we have not only relativistic bulk motion but also relativistic internal thermodynamic motion of particles constituting a plasma.

Importantly, the observed $\beta=1$ and the obtained $\Gamma=4/3$ indicate that the longitudinal current, though in general not obliged to conserve along the magnetic tube, appears as an exact integral of motion at the scales under consideration,
\begin{equation}
\label{longitudinalCurrentConservation}
I=\text{const}.
\end{equation}
Since $\rho\propto r^{-2-\beta}=r^{-3}$, we have for a hot flow $h\propto\rho^{\Gamma-1}=\rho^{1/3}\propto r^{-1}$; therefore, by Eq.~\eqref{powerLawAcceleration} the equality \eqref{gammaHConst} holds, whatever the nature of the force balance in the jet. Energy conservation \eqref{GSE} then leads us to the above conclusion. This is direct evidence of a stable longitudinal current structure in the jet.

It is interesting to consider what happens at $8$~marcsec. \citet{MertensEtal2016}, taking into account the results of \citet{Lyubarsky2009}, discuss that the jet may there enter a non-equilibrium regime or rather exhibit an early saturation of Poynting-flux conversion. Meanwhile, an alternative explanation may be a change of the jet thermodynamics. The relativistic gas is not characterized by a single polytropic index, and the general expression for its enthalpy is $h=K_3(1/\Theta)/K_2(1/\Theta)$, where $K_\nu(z)$ is the Macdonald function of order $\nu$ and $\Theta=p/\rho$ is the temperature \citep{Synge1957}. Useful simpler analytic approximations of the exact expression are presented in \citet{Mathews1971}, \citet{MelianiEtal2004}, \citet{ChattopadhyayRyu2009}. In the two opposite limiting cases it corresponds to a polytrope with index $\Gamma=4/3$ when $\Theta\gg1$ and with index $\Gamma=5/3$ when $\Theta\ll1$. From Eqs.~\eqref{powerLawAcceleration} and \eqref{powerLawAccelerationLargeZ} we have
\begin{equation}
\label{largeZ}
\beta_\text{obs}\approx\frac{0.16}{0.58}\approx0.28,\text{ }z>8\text{ marcsec},
\end{equation}
which allows us to formally estimate the polytropic index, respectively, for the case of a dominating longitudinal magnetic, transverse electromagnetic, and centrifugal pressure:
\begin{equation}
\label{polytropicIndexLargeZ}
\Gamma_\text{LM}\approx1.76,\text{ }\Gamma_\text{TEM}=\Gamma_\text{CH}\approx1.12,\text{ }\Gamma_\text{CC}\approx1.64.
\end{equation}
Since $\Gamma_\text{TEM}=\Gamma_\text{CH}<4/3$, a hot flow with a dominating transverse electromagnetic or centrifugal pressure is not realized over $8$~marcsec. The two other indices correspond to a cold flow: $\Gamma_\text{LM}$ is close to but yet slightly larger than the maximum possible $5/3$, while, surprisingly, $\Gamma_\text{CC}$ almost coincides with $5/3\approx1.67$. Therefore, a transition from $\gamma\propto z^{0.58}$ to $\gamma\propto z^{0.16}$ at $z\sim8$~marcsec may be explained by the plasma cooling and transition from a hot flow with $p\propto\rho^{4/3}$ to a cold flow with $p\propto\rho^{5/3}$. Over $8$~marcsec we likely have a cold flow whose pressure is determined mainly by centrifugal forces, possibly with some contribution from the pressure of longitudinal magnetic field, while transverse electromagnetic fields give forces that almost compensate each other and result in a modest net effect.

Determination of $\Gamma$ from the jet acceleration and collimation profiles has power in revealing plasma composition, which is important for conclusions about exact mechanisms of jet launching. The transition directly from $4/3$ to $5/3$ as cooling occurs may be a sign of a pure electron-positron plasma component: electrons and positrons have the same mass and behave as the mentioned single-component perfect Synge gas. In the case of an electron-proton plasma, however, the transition from $4/3$ to $5/3$ should occur through an intermediate value of $\Gamma\approx1.44$ taking place in a transitional temperature region and playing the role of an extra step at which polytropic index stays as dimensionless inverse temperature runs through the range $\sim10-100$ (see Fig.~1 in \citet{FalleKomissarov1996} and Figs.~1 and 2 in \citet{ChoiWiita2010}).

If we have a probable dominance of centrifugal forces not only above but also in the transitional region $z\sim8$~marcsec, we would observe in some range of distances about $8$~marcsec an acceleration profile of $\gamma\propto r^{0.46}\propto z^{0.27}$, see Eq.~\eqref{Gamma3ColdGeneral}, if we had a delay on $\Gamma\approx1.44$ during an electron-proton plasma cooling (we have taken into account that the collimation profile does not change at these distances). The observations by \citet{MertensEtal2016} do not show such an acceleration profile, so the existence of an electron-positron plasma component in the jet should be assumed. This is an evidence of the Blandford-Znajek jet-launching mechanism producing electron-positron pairs near the central engine \citep{BlandfordZnajek1977}.

Note that this conclusion does not mean that the entire jet is powered solely by the Blandford-Znajek mechanism, because there is evidence that the outer jet is likely produced via the Blandford-Payne mechanism. The radius of the jet base is $\sim10r_\text{g}$ \citep{MertensEtal2016}, which implies that the outer jet is launched from the accretion disc  \citep{BlandfordPayne1982}. Another evidence is the resolved triple-ridge structure of the jet \citep{AsadaEtal2016,Hada2017}, which can mean that the M87 jet is in fact jet in jet \citep{Sobyanin2017}, and different launching mechanisms for the inner and outer jets are not forbidden. Extra evidence for the Blandford-Payne mechanism goes from the observed quasi-periodic side-shift of the jet \citep{WalkerEtal2018}, which can be interpreted as jet precession resulting from Lense-Thirring precession of a tilted accretion disc, to which the jet is perpendicular \citep{Sobyanin2018}.

Most recent numerical simulations of tearing of a highly inclined accretion disc around a rotating black hole show that central parts of the disc undergo the Bardeen-Peterson effect and align with the black hole spin, and the radius of alignment is $<5-10r_\text{g}$ \citep{LiskaEtal2019}. The jet base radius then cannot be less than the alignment radius because otherwise precession would not be observed. This is consistent with the mentioned base radius estimate and possibility of the jet precession, favouring the Blandford-Payne launching mechanism for the outer jet. The inner jet, even if launched from the black hole, will align with the outer jet and be engaged in the joint precession because the transverse electromagnetic forces induce alignment as the inner jet is displaced from the axis of the outer jet \citep{Sobyanin2017}. This is consistent with the strict current conservation obtained above: the inner and outer jets, though bearing opposite charges and oppositely directed electric currents, do not touch each other and do not become short-circuited during precession.

None the less, the possibility of solely black-hole launching the entire jet is also discussed \citep{NakamuraEtal2018,OgiharaEtal2019}, and a direct observational sign of the electron-proton plasma composition would be helpful in ultimately distinguishing between the mixed Blandford-Znajek--Blandford-Payne and the pure Blandford-Znajek cases. The mentioned dependence $\gamma\propto r^{0.46}$ near $8$~marcsec (2.5~marcsec in projection) could be such a sign, and it is promising to observationally study the acceleration profile separately for the most outer areas of the jet at subpc scales with high longitudinal resolution.




\providecommand{\noopsort}[1]{}\providecommand{\singleletter}[1]{#1}%


\bsp	
\label{lastpage}
\end{document}